# Binning based algorithm for Pitch Detection in Hindustani Classical Music

Malvika Singh, BTech 4th year, *DAIICT,* 201401428@daiict.ac.in

*Abstract*—Speech coding forms a crucial element in speech communications. An important area concerning it lies in feature extraction which can be used for analyzing Hindustani Classical Music. An important feature in this respect is the fundamental frequency often referred to as the pitch. In this work, the terms pitch and its acoustical sensation, the frequency is used interchangeably. There exists numerous pitch detection algorithms which detect the main/ fundamental frequency in a given musical piece, but we have come up with a unique algorithm for pitch detection using the binning method as described in the paper using appropriate bin size. In particular, the paper [1] provides light on pitch identification for Hindustani Classical Music. Pitch Class Distribution has been employed in this work. It can be used to identify pitches in Hindustani Classical Music which is based on suitable intonations and swaras. It follows a particular ratio pattern which is a tuning for diatonic scale proposed by Ptolemy [8] and confirmed by Zarlino [9] is explored in this paper. We have also given our estimated of these ratios and compared the error with the above. The error produced by varying the bin size in our algorithm is investigated and an estimate for an appropriate bin size is suggested and tested. The binning algorithm thus helps to segregate the important pitches in a given musical piece.

## I. INTRODUCTION

Raga is a fundamental idea central to Hindustani Classical Music. They express characteristic moods and express various emotions in the way they are sung and performed. A veteran Hindustani Classical Vocalist has the perfect acoustical sense of timing, duration and frequency of the swaras and a high tonal quality, even though this art form is not documented and has been transferred through generations by oral recitation. In work [8], the authors have exhaustively discussed throughout the length and breadth of the paper the pitch detection for Hindustani Classical Music. However, this method utilizes pitch extractor, Pitch-class Distributions (PCDs) and Pitch-class Dyad Distributions (PCDDs) which have been used as features for raga identification. Our present work is based on developing a binning based algorithm which helps to identify the important pitches in Hindustani Classical Music which can be used to find a few interesting songs among the millions available in music retrieval systems for Hindustani Classical Music. As a consequence of the easy access to music, the field of music information retrieval (MIR) has emerged. Various pitch detection algorithms exists like autocorrelation method [1], HPS [2], RAPT [3], AMDF method [4], CPD [5], SIFT [6], DFE [7]. Although the algorithms are effectively robust, yet their accuracy deteriorates as the noise component increases. Pitch detection algorithms can be classified into the following categories:

1) Time domain
2) Frequency domain
3) Time and frequency domain jointly

In this paper, we discuss the method to identify the important pitches in Hindustani Classical Music by employing the binning method which helps in the process provided an accurate estimate of the bin size is made. It uses time and frequency domain jointly.

## II. MOTIVATION FOR DEVELOPMENT OF THE ALGORITHM

As demonstrated in [1], the paper explores various pitch class distribution and intonations for Hindustani Classical Music. It employs raga identification and performs pitch extraction using polyphonic melody extractor and obtaining pitch contours. In our experiment, however, we use the binning method to extract the pitches important in Hindustani Classical Music. In the research work of [1], it has been noticed in various performances like those of Pandit Bhimsen Joshi, one of the very notable artistes, Hindustani Classical Music vocalists in general are meticulous about the particular position on the music scale in which they intone a specific swara within its pitch interval. They have also observed and verified that these positions on the music scale are such that their frequencies are in ratios of small integers. It is important that the artiste adjusts his tone and swara so that it is in consonance in a musical performance with respect to the sequence of the notes in a particular musical piece. Professional performers would closely stick to these ratios. In work of [9], the variation in the frequencies of each swara for many ragas has been shown. In our work, we have calculated the these frequency ratios and compared them to the already existent Ptolemy and Gioseffo Zarlinos ratios along with the errors as depicted in Table 1 on the last page.

## III. ALGORITHM

The binning method focused on separating frequencies on the basis of how frequently they occurred during the music piece. This was done recursively by creating bins of appropriate sizes which then shortlisted the most frequently occurring frequency n the bin by comparison of frequencies within the bin. In this manner, frequencies were obtained locally which were then merged with similar frequencies to finally give the top ten frequencies. Here, care had to be taken so that we ignore very small frequencies outside the audible range even though they were the local maximums.

Finding the top 10 frequencies: In the first attempt, findPeaks method in MATLAB was used to find the peaks in the fft of the signal and the audio files. In this method, the peaks based on the amplitude magnitude were identified. This gave the frequencies which were very close,like, for example, for a given audio file, 168 and 168.2 Hz etc. which would practically be indistinguishable by the normal adult human ear. So, the need for an appropriate bin size was felt so as to group these closely related frequencies.

Figure 1 shows the graph plotted based on the frequency distribution of the signal, and so the top ten peaks can be manually found by visual inspection by observing the plot on looking at the number of times a particular frequency appears in the whole audio recording.

performed. Hence, we considered the *komal rishabh* and the ratio comes out to be 15/16 for it. In the above method, bins of an appropriate (described later), window size were made, and then, as soon as a frequency lying in that particular frequency appeared, we had to increment the bin count by 1 unit. The bin count was initially initialized with zeros. In this way we could find the top ten windows. Since the number of samples of the audio files was huge, in order to increase efficiency of the algorithm, we had to appropriately scale the signal. The bin size for classical Hindustani MUsic will depend on the frequency distribution and the ratios involved in this form of music. The following section describes as to why have we considered ratios and frequency distribution in order to figure out a good estimate of the bin size to be used in our algorithm.

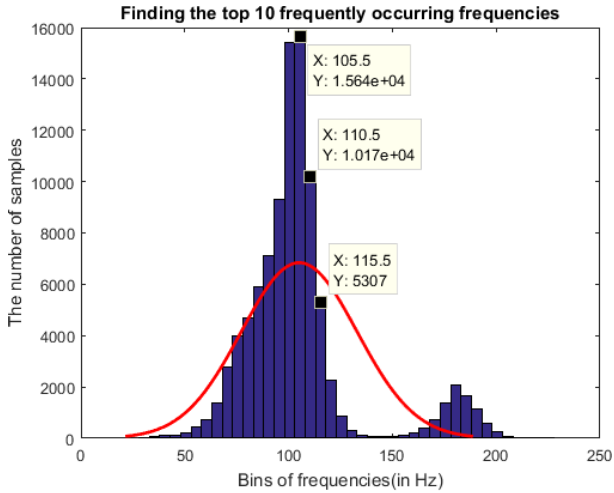

Fig 1. Probability distribution of top 10 frequencies

*A. Bin Size Estimation*

The bin size needs to be chosen in accordance to the inclination of most of the frequencies in the top ten frequency spectrum obtained earlier. The factors will vary depending upon the position of the majority of the frequencies falling as per the spectrum. The music scale can have the following divisions as per Hindustani Classical Music. Figure below illustrates these ratios:

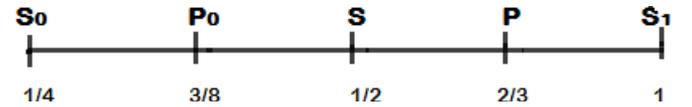

Fig 1. Music Scale in Hindustani Classical Music with ratios

**Algorithm 1** Binning algorithm

```
0: procedure BINMETHOD
0:    music ← array of audiodata
0:    fs ← sampling frequency of audiodata
0:    freq ← fft(music)
0:    binsize ← EstimateBinSize
0:    cnt ← 0
0:    distribution ← ProbabilityDistribution(freq)
0:    top:
0:    for i ← 1 to length(distribution) do
0:       bin[cnt] ← freq[i] to freq[i + binsize]
0:       Find maximum(bin_i)
0:       if maximum(bin_i threshold) then
0:          accept cnt ← cnt + 1
0:       else
0:          reject
0:       end if
0:    end for
0: end procedure=0
```

In the algorithm described above, bin size of 2Hz is taken into account. However, if the bin size is too small then the computation suffers and if it is too large then we miss out on the fundamental frequency. So a tradeoff needs to be

Now, we performed the experiment by taking into consideration each of the ratio in the above music scale and used the bin size estimate as 15/16 * Maximum Frequency*factor chosen. We plotted the original spectrum as below and the drone frequency obtained by using the appropriate bin size(the factor which gave the best estimate) was the factor where maximum frequencies lay in the original spectrum. Our dataset included classical Hindustani music like *raag behagra, Meera Kabir Bhajan, Raag Puria Kalyan* etc. Here are the results of the samples and the percentage error of the drone frequency with the original frequency obtained by xcorr function in MATLAB:

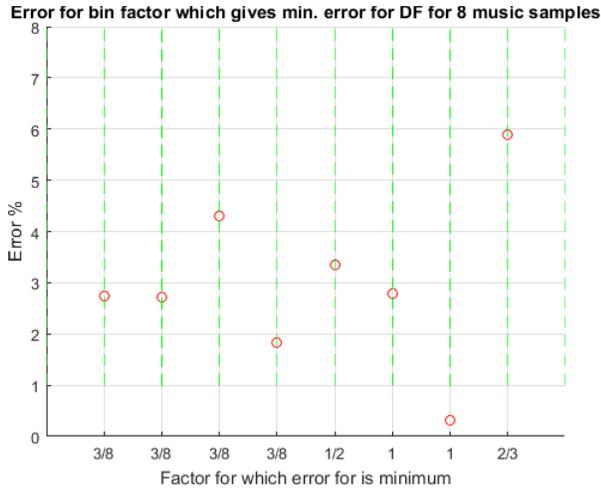

**Fig 2.** Error Percentage for bin estimate factor which gives minimum error for drone frequency for 8 music samples

As indicated by the plot above, the minimum error percentages are for those factors where the maximum number of frequency samples lie. The following 8 plots demostrate the frequency spectrums of which the error is plotted in order.

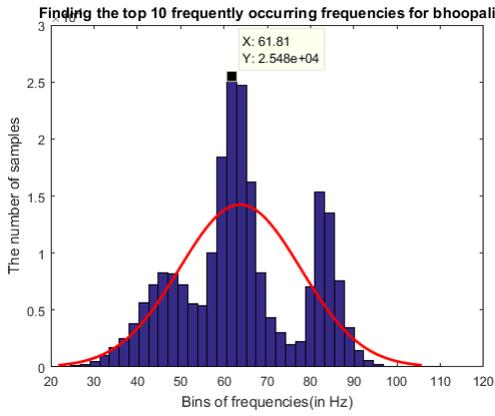

**Fig 3.** Probability distribution of frequencies for sample1

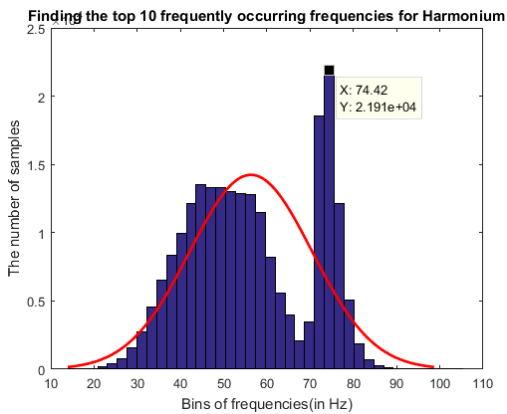

**Fig 4.** Probability distribution of frequencies for sample2

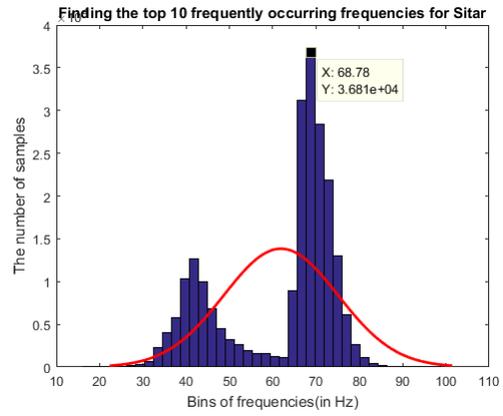

**Fig 5.** Probability distribution of frequencies for sample3

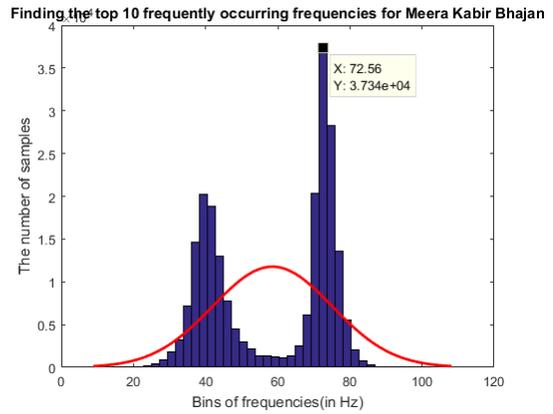

**Fig 6.** Probability distribution of frequencies for sample4

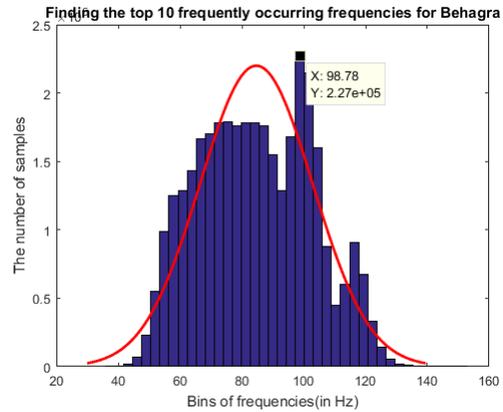

**Fig 7.** Probability distribution of frequencies for sample5

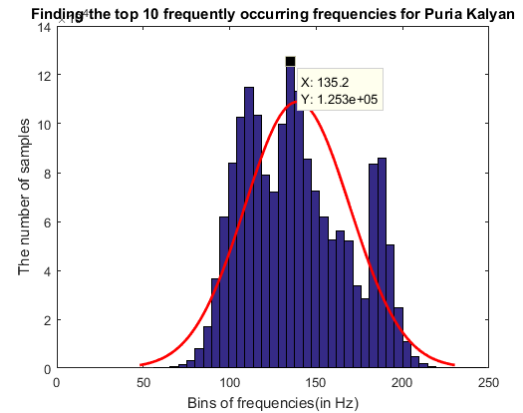

**Fig 8.** Probability distribution of frequencies for sample6

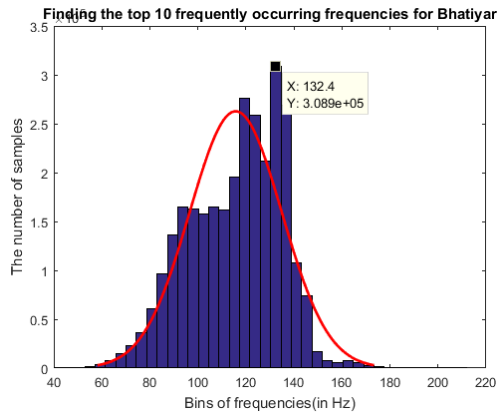

**Fig 9.** Probability distribution of frequencies for sample7

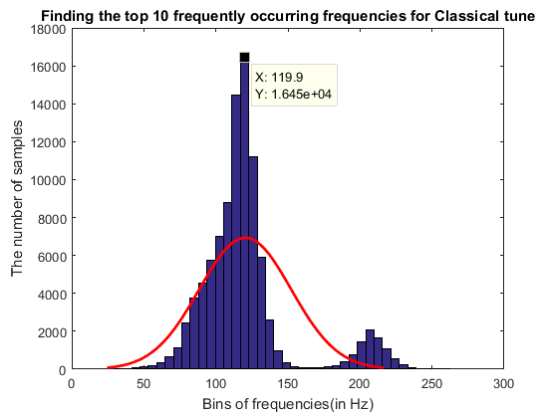

**Fig 10.** Probability distribution of frequencies for sample8

## IV. COMPARISON OF ERROR WITH DIFFERENT BIN SIZES

The experiment was performed to evaluate the error in binning algorithm when different bin sizes were considered. The experiment was performed using the classical Hindustani ragas *bhoopali*, music on the harmonium, the sitar and the *Meera-Kabir Bhajan* as well as some noise-free sound signal with the following specifications generated in MATLAB:
partial1 = cos(omega1*t + phi)*amplitude; %sinusoidal 1
partial2 = cos(omega2*t + phi)*amplitude; %sinusoidal 2
partial3 = cos(omega3*t + phi)*amplitude; %sinusoidal 3
signal = (partial1 + partial2 + partial3)/3; and varying frequencies between 100Hz to 500 Hz to generate different sounds. Figure 3 below shows the observation made in terms of the error percentage:

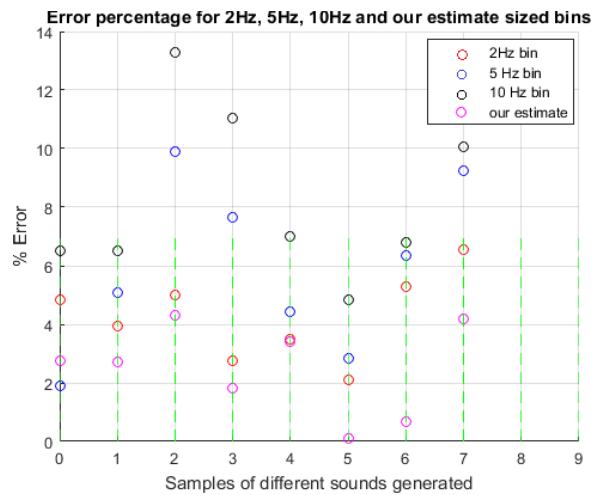

**Fig 11.** Comparing error percentage by varying bin size for sounds generated in MATLAB

As per the results obtained, the 2Hz bin size provides better performance than the 5Hz and 10Hz bins because the frequency taken into consideration is high, ranging from 100Hz to 500Hz. So the bin size has to be small so as to be more sensitive to frequency changes.

## V. CONCLUSION

Thus, in this work, we introduced a new method for obtaining the pitch of a music sound of moderate frequencies using matlab generated sound signals. We further observed the effect of varying the bin size for these sound signals. In this process, we provided an estimate of an appropriate bin size. In addition to this, the error analysis shows the percentage error when bin size is varied for different sound signals. This helps in identifying the ragas in Hindustani Classical Music which is unique in its own way. It does not have accurate documented musical notes like in case of Western Music, rather every artiste adjusts the intonation slightly to suit his/her voice while adhering to the ratios, but via an acoustical feel and not a precise measurement. The work contributes to get a measurement for the fundamental frequency so that learning of this form of music can be learnt even by novice musicians by knowing the frequency they need to achieve along with the tacit

### ACKNOWLEDGMENT

The authors would like to mention Miss Ahana Pradhan, PHD, IITBombay and Miss Sobha Singh, MTech IIT-Bombay for their invaluable suggestions in the process.

| Musical Note | Ptolemy's ratios | Our calculation of ratios | Gioseffo Zarlinos ratios | Diff.(Ptolemy& our ratios) | Diff.(Zarlino&our ratio) |
|---|---|---|---|---|---|
| C | 1/1 | 1/1 | 1/1 | 0 | 0 |
| C# | 16/15 | 256/243 | 25/24 | 0.01316 | 0.01183 |
| B | 15/8 | 243/128 | 15/8 | 0.02343 | 0.02343 |
| B# | 9/5 | 9/5 | 9/5 | 0 | 0 |
| E | 5/4 | 81/64 | 5/4 | 0.01562 | 0.01562 |
| F | 4/3 | 4/3 | 4/3 | 0 | 0 |
| F# | 45/32 | 729/512 | 45/32 | 0.01757 | 0.01757 |
| G | 3/2 | 3/2 | 3/2 | 0 | 0 |
| G# | 8/5 | 128/81 | 25/16 | 0.01975 | 0.01774 |
| A | 5/3 | 27/16 | 5/3 | 0.02083 | 0.02083 |
| A# | 9/5 | 16/9 | 16/9 | 0.02222 | 0.02222 |
| C | 2/1 | 2/1 | 2/1 | 0 | 0 |

TABLE I

**The ratios for different tones from Ptolemy's work, Gioseffo Zarlino's work and our calculations**